\documentclass[aps,prl,
showpacs,superscriptaddress,groupedaddress]{revtex4-1}  
\usepackage{graphicx}  
\usepackage{dcolumn}   
\usepackage{bm}        
\usepackage{amssymb}   
\usepackage{amsmath}
\DeclareMathOperator{\arccot}{arccot}

\begin{document}



\title{Fano Interference in Two-Photon Transport}
\author{Shanshan Xu}
 \affiliation{Department of Physics,
Stanford University, Stanford, California 94305}

\author{Shanhui Fan}
\email{shanhui@stanford.edu} \affiliation{Department of Electrical
Engineering, Ginzton Laboratory, Stanford University, Stanford,
California 94305}


\begin{abstract}
We present a general input-output formalism for the few-photon transport in multiple waveguide channels coupled to a local cavity.
Using this formalism, we study the effect of Fano interference in two-photon quantum transport. We show that the physics of Fano
interference can manifest as an asymmetric spectral line shape in the frequency dependence of the two-photon correlation function. The 
two-photon fluorescence spectrum, on the other hand, does not exhibit the physics of Fano interference.
 \end{abstract}

\maketitle

\section{\label{sec:level1} I. Introduction}

Fano resonance was originally discovered in the study of atomic physics \cite{fano} and manifests as an asymmetric line shape in the atomic absorption spectrum. In recent years, however,
there  has been an explosion of interest in exploration of Fano interference effect in nanophotonic structures \cite{mfk,lzm,f,fsj,mmfk,mk,mmkb,hsdmhn,llhn,brbjffb,zgbr,bcpbp}. Such interest arises since Fano interference effects in fact
occur rather frequently in a wide variety of nanophotonic  structures. Moreover, there has been strong interest to use Fano interference effect for applications such as 
add-drop filters \cite{fvjh,fvjkmh,sljf}, sensors \cite{hsdmhn,hnsdm,llmw} and optimal bistable switches \cite{cy,mbb,mmk,yhwyk}. 

In nanophotonics, the Fano interference effect is typically a linear effect where the interference results in an asymmetric line shape in the transmission and reflection spectra. On the other hand,
there has been interest in exploiting such linear interference effect as a basis for engineering and enhancing nonlinear optical interactions, in particular for optical switching applications \cite{mmfk,mk,mmkb,yfs,aem,ntsmstn,hkem}.
In all these studies, light is treated as classical electromagnetic waves. Connected to these works, it would be interesting to explore the implication of Fano interference effect in the quantum regime for photon-photon interaction.
The effect of Fano interference in single photon transport has been studied in \cite{jsfano}.
In the study of waveguide quantum electrodynamics system, there has also been extensive studies on multi-photon transport \cite{sf,js,fks,zgb2,lsb,eks,zb,r,koz,lhl,sek,rwf,smzg,ll,ll2,sfs,rf,jg,lnsa,srf,xf,cmsdcc,scc,rs,xl}. 
However, there has not been any study on the effect of single-photon Fano interference for the quantum transport of multi-photon Fock state, especially in the presence of strong photon-photon interactions.

In this paper, we consider a general waveguide quantum electrodynamics (QED) system of a localized optical mode, such as those found in an optical cavity, coupling to multiple input and output 
waveguide channels. The cavity in addition can incorporate nonlinear elements that result in strong optical nonlinearity at the two-photon level such as Kerr-nonlinear cavity or optomechanical cavity\cite{ll2}. For this system, we develop an
 input-output formalism to study the effect of Fano resonance in the single and two-photon transport. We show that 
the coupling constants in such formalism are strongly constrained by the fundamental principles of quantum mechanics and symmetry considerations. For single photon transport,
Fano line shape directly arises from these constraints. 
For two-photon transport, we show that the Fano line shape does not appear in the two-photon resonance fluorescences. Nevertheless, an asymmetric line shape related to
Fano interference does appear in two-photon correlation functions. 

The paper is organized as follows. 
In Section II we present a general input-output formalism of Fano resonance in a local cavity coupled to multiple waveguide channels. In Section III  we prove that the coupling constants are 
strongly constrained by fundamental principles of quantum mechanics and time-reversal symmetry. 
 In Section IV we provide a general formula of the multi-photon S matrix and the resulting two-photon correlation functions based on the input-output formalism. 
 In Section V,  we consider a specific example of a two-mode waveguide coupled to a two-level atom that exhibits Fano resonance in the single photon S matrix and 
 two-photon correlation functions. 

\section{II. Input-Output Formalism}
We consider a general class of waveguide QED systems that consist of a cavity coupled to $N$ 
waveguide channels. For simplicity we assume that the cavity has only one mode, as described by the bosonic operator $a$, 
that couples to the waveguide channels. $[a, a^{\dag}]=1$. Also, for our purpose here, the cavity can support strong nonlinearity at the few photon level. 
The waveguide channels can arise from the use of multiple waveguides, or they can correspond to modes of a single multi-mode waveguide. 
Also, in consistency with the waveguide QED literature \cite{js}, modes propagating along forward and backward directions are treated as different channels.
The dynamics of the system is described by the input-output formalism \cite{gc, fks, xf}:
\begin{eqnarray}
\frac{d}{dt}\,a&=&-i\left[a, H_{\text{c}}\right]-\,\Sigma\,a+\,\boldsymbol{\kappa}^T\,\mathbf{ c}_{\text{in}}\,\label{io3}\\
\mathbf{ c}_{\text{out}}(t)&=&\mathbf{C}\,\mathbf{ c}_{\text{in}}(t)+\,a(t)\,\mathbf{d}\,,\label{io1}\label{io1}
\end{eqnarray}
where $H_{\text{c}}=H^{\dag}_{\text{c}}$ is the Hamiltonian of the cavity. $\Sigma$  is the self-energy correction due to the coupling of the cavity to the waveguide channels. 
The imaginary part of $\Sigma$ describes the decay of the cavity mode. The resonance is excited by the input operators 
\begin{equation}
\mathbf{ c}_{\text{in}}(t)=\left[\begin{array}{c} c_{1,\text{in}}(t)\\c_{2,\text{in}}(t)\\ \vdots \\c_{N,\text{in}}(t)\end{array}\right]
\end{equation} 
from waveguide channels $1$ to $N$, respectively, with the coupling constants
\begin{equation}
\boldsymbol{\kappa}=\left[\begin{array}{c} \kappa_1\\ \kappa_2\\ \vdots \\\kappa_N\end{array}\right]\,.
\end{equation}
The resonant excitation can decay into the waveguides and couples with the output operators 
\begin{equation}
\mathbf{ c}_{\text{out}}(t)=\left[\begin{array}{c} c_{1,\text{out}}(t)\\c_{2,\text{out}}(t)\\ \vdots \\c_{N,\text{out}}(t)\end{array}\right]
\end{equation}
with the coupling constants
\begin{equation}
\mathbf{ d}=\left[\begin{array}{c} d_1\\d_2\\ \vdots \\d_N\end{array}\right]\,.
\end{equation}
As seen in (\ref{io1}), the excitation of the resonance therefore results in the coupling between the input and output of different waveguide channels. 
In addition to such a resonant pathway for scattering among different waveguide channels, the channels may also couple among themselves directly in the absence of the resonance, 
as defined by a scattering matrix $\mathbf{C}$ that describes such a coupling. The scattering matrix $\mathbf{C}$ for the direct pathway is unitary, i.e.
 $\mathbf{C}\mathbf{C}^{\dag}=\mathbf{I}$.  In the standard input-output formalism \cite{gc, fks, xf}, $\mathbf{C} = \mathbf{I}$.  Here we assume that $\mathbf{C}$ is an arbitrary unitary matrix in order to treat Fano interference. 
 
In Appendix A.1, we provide the underlying Hamiltonian of a specific  waveguide QED system from which one can derive the input-output formalism that has the form of (\ref{io3})-(\ref{io1}). 
As far as the scattering properties are concerned, however, the input-output formalism provides a more convenient starting point for the theoretical developments. 
Therefore, in consistency with much of the quantum optics literature where the input-output formalisms are used, one can in fact treat (\ref{io3}) and (\ref{io1}) as the starting ansatz. 
Also, (\ref{io3}) and (\ref{io1}) are similar to that in the temporal coupled mode theory widely used in nanophotonics to treat classical electromagnetic effects, 
including the Fano interference effects \cite{fsj}, but here all the dynamic variables are Heisenberg operators rather than c-numbers.

\section{III. Constraints on the coupling constants}
For a given scattering matrix $\mathbf{C}$ in the direct pathway, 
the coupling constants $\boldsymbol{\kappa}$, $\mathbf{d}$ and self-energy correction $\Sigma$ in the input-output formalism (\ref{io3})-(\ref{io1}) cannot be arbitrary. 
Instead, they are related to one another by fundamental principles in quantum mechanics and additional symmetry requirements. 
In this section, we derive the constraints imposed 
by flux conservation, quantum causality and time-reversal symmetry.

\subsection{A. Flux Conservation}

We suppose that the cavity system, in the absence of the waveguide, conserves the total number of excitations inside the cavity, i.e. there exists a conserved excitation number operator $N$ for the total number of excitations, satisfying 
\begin{equation}\label{NH}
\left[N,\,H_c\right]=0\,.
\end{equation}
The operator $N$ takes non-negative integer as its eigenvalues. Removing a cavity photon should reduce the total number of excitations in the cavity system by unity, and hence
\begin{equation}\label{oN}
\left[N,\,a\right]=-a\,.
\end{equation}
In order to satisfy the commutation relation (\ref{oN}), a natural form of the number operator $N$ is therefore
\begin{equation}\label{tN}
N=a^{\dag} a+O\,,
\end{equation}
where $O$ consists of other degrees of freedom in the cavity with $\left[a,\,O\right]=0$. In our form of the input-output formalism (\ref{io3})-(\ref{io1}), only the cavity operator $a$ couples to the waveguide, whereas these other degrees of freedom do not couple with the waveguide directly, i.e.
\begin{equation} \label{addo}
\frac{d}{dt}\,O=-i\,\left[O,\,H_c\right]\,.
\end{equation}

Having defined the excitation number operator, we can describe the condition for flux conservation as:
\begin{equation}\label{fluxio}
\frac{d}{dt}\,N= \mathbf{ c}_{\text{in}}^{\dag}\,\mathbf{ c}_{\text{in}}-\mathbf{ c}^{\dag}_{\text{out}}\,\mathbf{ c}_{\text{out}}\,.
\end{equation}
Using (\ref{io3}), as well as (\ref{NH})-(\ref{addo}) that describe the properties of the excitation number operator, we have
\begin{equation}\label{fccc}
\frac{d}{dt}\,N= -\left(\Sigma+\Sigma^*\right)\, a^{\dag}a+ \mathbf{ c}_{\text{in}}^{\dag} \,\boldsymbol{\kappa}^* \,a+a^{\dag} \,\boldsymbol{\kappa}^T\,\mathbf{ c}_{\text{in}}\,.
\end{equation}
On the other hand, using (\ref{io1}) we have
\begin{equation}
\mathbf{ c}_{\text{in}}^{\dag}\,\mathbf{ c}_{\text{in}}-\mathbf{ c}^{\dag}_{\text{out}}\,\mathbf{ c}_{\text{out}}
= - \mathbf{d}^{\dag} \mathbf{d}\, a^{\dag}a- \mathbf{ c}_{\text{in}}^{\dag} \,\mathbf{C}^{\dag}\,\mathbf{d} \,a-a^{\dag} \,\mathbf{d}^{\dag}\,\mathbf{C}\,\mathbf{ c}_{\text{in}}\,.
\end{equation}
The flux conservation condition (\ref{fccc}) then requires
\begin{equation}\label{fluxc}
\mathbf{d}^{\dag} \mathbf{d}=\Sigma+\Sigma^*\,,\,\,\,\,\,\,\,\, \mathbf{C}^{\dag}\,\mathbf{d}=-\boldsymbol{\kappa}^*\,.
\end{equation}

\subsection{B. Quantum Causality}

In the input-output formalism (\ref{io3})-(\ref{io1}), the operator $a(t)$, which characterizes the physical field in the cavity, depends only on the input field $\mathbf{c}_{\text{in}}(t')$ with  $t'<t$, and generates only output field
$\mathbf{c}_{\text{out}}(t')$ with $t'> t$. This can be formulated as the quantum causality condition \cite{gc,xf}:
\begin{eqnarray}
\left[a(t)\,,\mathbf{c}^{\dag}_{\text{in}}(t')\right]&=&0\,,\,\,\,\,\,\,\,\,\text{for}\,\,t<t'\label{fff1}\\
\left[a(t)\,,\mathbf{c}^{\dag}_{\text{out}}(t')\right]&=&0\,,\,\,\,\,\,\,\,\,\text{for}\,\,t>t'\,.
\end{eqnarray}
The commutator $\left[a(t)\,,\mathbf{c}^{\dag}_{\text{in}}(t')\right]$ for $t>t'$ can then be computed as:
\begin{eqnarray}
\left[a(t)\,,\mathbf{c}^{\dag}_{\text{in}}(t')\right]&=&\left[a(t)\,,\mathbf{c}^{\dag}_{\text{out}}(t')\mathbf{C}-a(t')\mathbf{d}^{\dag}\mathbf{C}\right]=-\mathbf{d}^{\dag}\mathbf{C}\,\left[a(t)\,,a^{\dag}(t')\right]\,.\label{fff2}
\end{eqnarray}
Combining (\ref{fff1}) and (\ref{fff2}) leads to the relation
\begin{eqnarray}
\left[a(t)\,,\mathbf{c}^{\dag}_{\text{in}}(t')\right]&=&-\mathbf{d}^{\dag}\mathbf{C}\,\left[a(t)\,,a^{\dag}(t')\right]\,\theta(t-t')\,,\label{qc1}
\end{eqnarray}
where
\begin{equation}
\theta(t)\equiv\left\{\begin{array}{c}1\,\,\,\,\,\,\,\,\,\,\,\,\,\,\,\,t>0    \\1/2\,\,\,\,\,\,\,\,\,\,t=0 \\0\,\,\,\,\,\,\,\,\,\,\,\,\,\,\,\,t<0\end{array}\right.\nonumber
\end{equation}
is the Heaviside step function. Similarly, we can derive
\begin{equation}
\left[a(t)\,,\mathbf{c}^{\dag}_{\text{out}}(t')\right]=\mathbf{d}^{\dag}\,\left[a(t)\,,a^{\dag}(t')\right]\,\theta(t'-t)\,.\label{qc2}
\end{equation}

Since $a$ and $a^{\dag}$ here are the bosonic creation and annihilation operators, we expect 
$\frac{d}{dt}\left[a(t), a^{\dag}(t)\right] $ to vanish for all $t$ because of the equal-time commutator 
$\left[a(t), a^{\dag}(t)\right]=1$. 
Therefore, using (\ref{io3}) we have 
\begin{eqnarray}
\frac{d}{dt}\left[a(t), a^{\dag}(t)\right]&=&-\left(\Sigma+\Sigma^*\right)[a(t), a^{\dag}(t)]+\boldsymbol{\kappa}^T\left[\mathbf{c}_{\text{in}}(t), a^{\dag}(t)\right]
+\left[a(t), \mathbf{c}^{\dag}_{\text{in}}(t)\right]\boldsymbol{\kappa}^*=0\,.
\end{eqnarray}
Further applying (\ref{qc1}) and (\ref{qc2}) leads to a constraint between $\mathbf{C}$, $\mathbf{d}$, $\boldsymbol{\kappa}$, and $\Sigma$:
\begin{equation}\label{constr1}
\Sigma+\Sigma^*=-\frac{1}{2}\left[\boldsymbol{\kappa}^T\,\mathbf{C}^{\dag}\,\mathbf{d}\,+\,\mathbf{d}^{\dag}\,\mathbf{C}\,\boldsymbol{\kappa}^*\right]\,.
\end{equation}
Note that this constraint is implied by the constraint of flux conservation (\ref{fluxc}).

\subsection{C. Time-Reversal Symmetry}
There are additional constraints on the parameters $\mathbf{C}$, $\mathbf{d}$, $\boldsymbol{\kappa}$, and $\Sigma$ from symmetries of the full system. 
Here we consider the implication of time-reversal symmetry. 
Let $\Theta$ be the time-reversal operator, which is antiunitary and in addition has the following properties:
\begin{equation}\label{pts}
\Theta\, \mathbf{c}_{\text{in}}(t)\, \Theta^{-1}=\mathbf{c}_{\text{out}}(-t)\,,
\,\,\,\,\,\,\,\Theta \,a(t) \,\Theta^{-1}=a(-t)\,,\,\,\,\,\,\,\,\Theta \,H_{\text{c}} \,\Theta^{-1}=H_{\text{c}}\,.
\end{equation} 
In Appendix A. 1, we provide a concrete construction of such an operator for a specific waveguide QED Hamiltonian.

 Under the operation of $\Theta$, the input-output formalism (\ref{io3})-(\ref{io1}) becomes:
\begin{eqnarray}
\frac{d}{dt}\,a&=&-i\left[a, H_{c}\right]-(\boldsymbol{\kappa}^{\dag}\mathbf{d}-\Sigma^*)\, a-\boldsymbol{\kappa}^{\dag}\,\mathbf{C}\,\mathbf{ c}_{\text{in}}\,.\label{trio3}\\
\mathbf{ c}_{\text{out}}(t)&=&\mathbf{C}^{T}\,\mathbf{ c}_{\text{in}}(t)-a(t)\,\mathbf{C}^{T}\,\mathbf{d}^*\,,\label{trio1}
\end{eqnarray}
On the other hand, suppose the system has time-reversal symmetry, (\ref{trio3}) and (\ref{trio1}) should be identical to (\ref{io3}) and (\ref{io1}), and consequently, we have
 \begin{eqnarray}\label{trs}
 &&\mathbf{C}=\mathbf{C}^{T}\,,\,\,\,\,\,\,\mathbf{C}\,\mathbf{d}^*=-\mathbf{d}\,,\,\,\,\,\,\,\mathbf{C}\,\boldsymbol{\kappa}^*=-\boldsymbol{\kappa}\,,\,\,\,\,\,\,\boldsymbol{\kappa}^{\dag}\mathbf{d}=\Sigma+\Sigma^*\,.
 \end{eqnarray} 

In the following we will focus on systems that have the properties of flux conservation, quantum causality and time-reversal symmetry. That is, the parameters in the input-output formalism (\ref{io3})-(\ref{io1})
satisfy all the constraints of  (\ref{fluxc}),  (\ref{constr1}) and (\ref{trs}). As a result, they are related to each other as
\begin{eqnarray}
&&\mathbf{C}\,\mathbf{C}^{\dag}=\mathbf{I}\,,\,\,\,\,\,\,\mathbf{C}=\mathbf{C}^{T}\,,\label{fcc}\\
&&\mathbf{C}\,\mathbf{d}^*=-\mathbf{d}\,,\label{fcc1}\\
&&\mathbf{d}^{\dag}\mathbf{d}=\Sigma+\Sigma^*\,,\label{fcc2}\\
&&\boldsymbol{\kappa}=\mathbf{d}\,.\label{fcc3}
\end{eqnarray}
In particular, $\boldsymbol{\kappa}$ is determined from $\mathbf{d}$.

We note that the input-output formalism (\ref{io3})-(\ref{io1}), as well as the constraints on the parameters as shown in (\ref{fcc})-(\ref{fcc3}), agrees in form with the temporal coupled mode theory as developed in Ref.\cite{fsj} for classical 
electromagnetic waves. 
In deriving (\ref{fcc})-(\ref{fcc3}), the constraints of flux conservation and time-reversal symmetry correspond to similar constraints used in the development of the temporal coupled mode theory \cite{fsj}. 
The quantum causality condition has no classical counter parts. However, as we have shown in section III B above, 
the quantum causality condition does not give any additional constraint if one assumes flux conservation. Our results here show that the temporal coupled mode theory that was previously developed for classical waves can be adopted to treat the propagations of quantized electromagnetic waves and can be applied for few-photon states.

\section{IV. Multi-photon S matrix}

The single and two-photon S matrices are defined by the input and output operators as \cite{fks,xf}
\begin{eqnarray}
S_{\mu p;\nu k}&=&\int \frac{dt'}{\sqrt{2\pi}}e^{ipt'}\int \frac{dt}{\sqrt{2\pi}}e^{-ikt}\langle 0|c_{\mu,\text{out}}(t')\,c^{\dag}_{\nu,\text{in}}(t)|0\rangle\,,\\
S_{\mu p_1,\nu p_2;\rho k_1, \sigma k_2}&=&\left(\prod_{l=1}^2\int \frac{dt_l'}{\sqrt{2\pi}}e^{ip_lt'_l}\int \frac{dt_l}{\sqrt{2\pi}}e^{-ik_lt_l}\right)\langle 0|c_{\mu,\text{out}}(t_1')c_{\nu,\text{out}}(t_2')\,c^{\dag}_{\rho,\text{in}}(t_1)c^{\dag}_{\sigma,\text{in}}(t_2)|0\rangle\,,
\end{eqnarray}
where $\mu, \nu, \rho, \sigma$ denote waveguide channels that take values from 1 to $N$. 
With the input-output formalism (\ref{io3})-(\ref{io1}) and the quantum causality condition (\ref{qc1})-(\ref{qc2}), we can adopt the computational scheme in 
\cite{xf} to decompose the S matrices as
\begin{eqnarray}
S_{\mu p;\nu k}&=&\mathbf{C}_{\mu\nu}\,\delta(p-k)+S^C_{\mu p;\nu k}\,,\label{s1}\\
S_{\mu p_1,\nu p_2;\rho k_1, \sigma k_2}&=&S_{\mu p_1;\rho k_1}S_{\nu p_2;\sigma k_2}+S_{\mu p_1;\sigma k_2}S_{\nu p_2;\rho k_1}+S^C_{\mu p_1,\nu p_2;\rho k_1, \sigma k_2}\,,\label{s2}
\end{eqnarray}
where $S^C_{\mu p;\nu k}$ and $S^C_{\mu p_1,\nu p_2;\rho k_1, \sigma k_2}$ are related to the connected parts of the cavity's Green functions:
\begin{eqnarray}
S^C_{\mu p;\nu k}&=&\mathbf{d}_{\mu}\mathbf{d}_{\nu}\int \frac{dt'}{\sqrt{2\pi}}e^{ipt'}\int \frac{dt}{\sqrt{2\pi}}e^{-ikt}\,\langle 0| {\cal{T}}a(t')a^{\dag}(t)|0\rangle\,,\label{SC1}\\
S^C_{\mu p_1,\nu p_2;\rho k_1, \sigma k_2}&=&\mathbf{d}_{\mu}\mathbf{d}_{\nu}\mathbf{d}_{\rho}\mathbf{d}_{\sigma}\left(\prod_{l=1}^2\int \frac{dt'_l}{\sqrt{2\pi}}e^{ip_lt'_l}\int \frac{dt_l}{\sqrt{2\pi}}e^{-ik_lt_l}\right)
\,\langle 0| {\cal{T}}a(t'_1)a(t'_2)a^{\dag}(t_1)a^{\dag}(t_2)|0\rangle_C\label{SC2}\,.
\end{eqnarray}
Using the flux conservation relation (\ref{fluxc}), we prove in Appendix A.2 that Green functions in (\ref{SC1}) and (\ref{SC2})  can be computed using the effective Hamiltonian of the cavity
\begin{equation}\label{effH}
H_{\text{eff}}=H_{\text{c}}-i\,\Sigma\, a^{\dag}a
\end{equation}
without involving any waveguide degrees of freedom.

The single photon S matrix (\ref{s1}) shows explicitly that the transmission amplitude of single photon transport comes from the interference between a directly pathway as described by the scattering matrix $\mathbf{C}$ and a resonance-assisted indirect pathway  
as described by the two-point Green function of the cavity. This interference can produce the Fano resonance in transmission spectrum \cite{fsj}.	

For the two-photon transport,  the two-photon resonance fluorescence spectrum is described by the connected two-photon S matrix as $\left|S^C_{\mu p_1,\nu p_2;\rho k_1, \sigma k_2}\right|^2$. From (\ref{SC2}), for different waveguide channels, 
 the two-photon fluorescence spectrums are the same up to an overall prefactor $\left|\mathbf{d}_{\mu}\mathbf{d}_{\nu}\mathbf{d}_{\rho}\mathbf{d}_{\sigma}\right|^2$. They 
 share the same pole structure  that is completely determined by the cavity's effective Hamiltonian (\ref{effH}). 
 Therefore,  
 we expect that two-photon resonance fluorescence spectrums are qualitatively the same as that in the single channel case and thus there are no Fano interferences in the fluorescence spectrums. 
 
 On the other hand, the full two-photon S matrix in (\ref{s2}) does contain contributions from the direct pathway ($S_{\mu p;\nu k}$ contains contribution from $\mathbf{C}$), as well as the resonant pathway related to the cavity excitation. Therefore,
 the physics of Fano interference should appear in experimental quantities that are directly related to the two-photon S matrix. As an example, we consider the two-photon correlation function that describes the photon statistics of the outgoing two-photon state. 
 Without loss of generality, we consider an incident two-photon planewave state $|\nu,k_1; \nu,k_2\rangle$, comprised of two photons in the waveguide channel $\nu$ with
individual frequencies $k_1$ and $k_2$, as described by
\begin{equation}
|\nu,k_1; \nu,k_2\rangle=\int dx_1dx_2 P_{k_1k_2}(x_1,x_2) \frac{1}{\sqrt{2}}c^{\dag}_{\nu}(x_1)c^{\dag}_{\nu}(x_2)|0\rangle\,,
\end{equation}  
where $P_{k_1k_2}(x_1,x_2)=\frac{1}{\sqrt{2}2\pi}\left[e^{ik_1x_1}e^{ik_2x_2}+e^{ik_1x_2}e^{ik_2x_1}\right]$ is a symmetrized two-photon planewave and  $c^{\dag}_{\mu}(x)$ is the creation operator in the coordinate space with the commutation relation 
$\left[c_{\mu}(x), c_{\nu}^{\dag}(y)\right]=\delta_{\mu,\nu}\delta(x-y)$.
Here we focus on the resulting outgoing state with both photons in the same waveguide channel, say the $\mu$-th channel. From the two-photon S matrix (\ref{s2}), the outgoing state is 
\begin{eqnarray}
 |\phi\rangle_{\mu\mu}
 &=&  \frac{1}{2}\int dp_1dp_2\, \left|\mu, p_1; \mu, p_2\right.\rangle\,S_{\mu p_1,\mu p_2;\nu k_1, \nu k_2} \nonumber\\
 &=&\int dx_1dx_2\left[t_{\mu\nu}(k_1)t_{\mu\nu}(k_2)P_{k_1k_2}(x_1,x_2)+H_{\mu\nu}(x_1,x_2)\right]\frac{1}{\sqrt{2}}c^{\dag}_{\mu}(x_1)c^{\dag}_{\mu}(x_2)|0\rangle\,,\label{phimm}
\end{eqnarray}
where $t_{\mu\nu}(k)$ is the single photon transmission amplitude that is defined by the single photon S matrix as $S_{\mu p;\nu k}\equiv t_{\mu\nu}(k)\delta(p-k)$.
$H_{\mu\nu}(x_1,x_2)$ is the wavefunction of the two-photon bound state that is determined by the connected two-photon S matrix as \cite{sf,zgb2}
\begin{equation}\label{Hmi}
H_{\mu\nu}(x_1,x_2)=\frac{1}{2}\int dp_1dp_2\,P_{p_1p_2}(x_1,x_2)\, S^C_{\mu p_1,\mu p_2;\nu k_1, \nu k_2}\,.
\end{equation}
The two-photon correlation function associated with the outgoing state (\ref{phimm}) can then be computed by 
\begin{eqnarray}
G^{(2)}(\tau)&=&\,_{\mu\mu}\langle \phi|c_{\mu}^{\dag}(y)c_{\mu}^{\dag}(y+\tau)c_{\mu}(y+\tau)c_{\mu}(y)|\phi\rangle_{\mu\mu} \nonumber\\
&=&2\left| t_{\mu\nu}(k_1)t_{\mu\nu}(k_2)P_{k_1k_2}(y+\tau,y)+H_{\mu\nu}(y+\tau,y) \right|^2\,.\label{G2y}
\end{eqnarray}

In  (\ref{G2y}), $t_{\mu\nu}(k_1)$ and $t_{\mu\nu}(k_2)$ are the single photon transmission amplitude, which in general can exhibit an asymmetric Fano line shape with respect to $k_1$ and $k_2$. Therefore, the two-photon
correlation function is also influenced by Fano interference.

\section{V. A two-mode waveguide coupled to a two-level atom}
\begin{figure}
\includegraphics[width=0.8\textwidth] {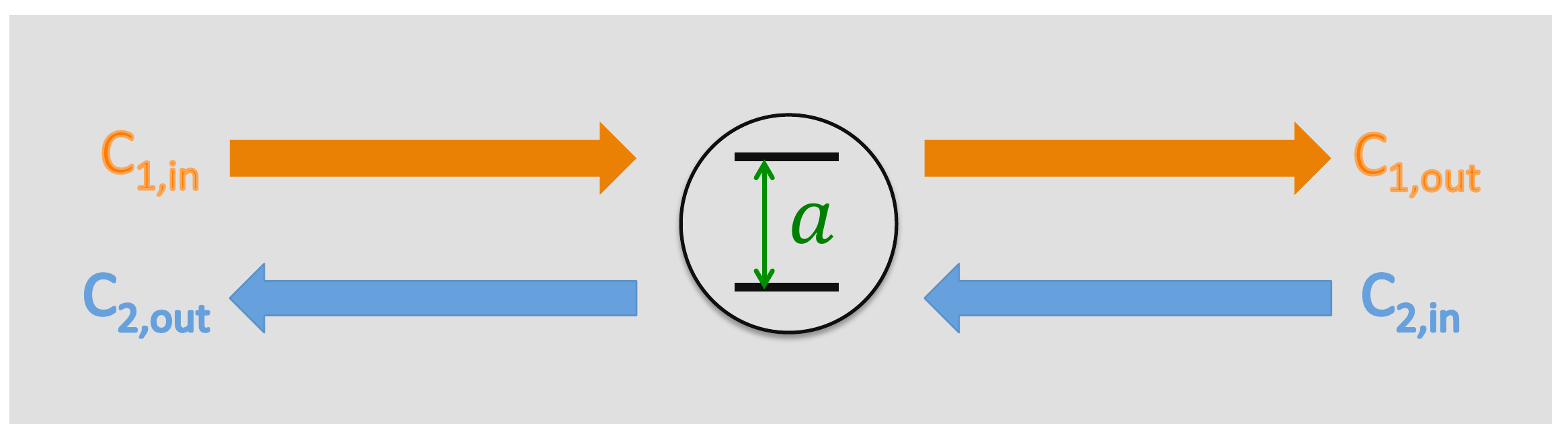}   \caption{
A two-mode waveguide coupled to a two-level atom.
} \label{w2ls}
\end{figure}

To support the general argument above, in this section we consider  a specific example of a two-mode waveguide coupled to a two-level atom. Each mode of the waveguide is treated as an individual channel
so that the transport property can be described by the input-output formalism (\ref{io3})-(\ref{io1}). 
As shown in Fig.\ref{w2ls},
the left-moving photon state is related to the operators $c_{1,\text{in}}(t)$ and  $c_{1,\text{out}}(t)$ while  $c_{2,\text{in}}(t)$ and  $c_{2,\text{out}}(t)$ represent the right-moving photon state. 

The two-level atom is described by
the Hamiltonian 
\begin{eqnarray}\label{Hatom}
H_{\text{atom}}=\Omega\, |e\rangle\langle e|\,,
\end{eqnarray}
where $\Omega$ is the atom's resonant frequency and $|g\rangle,\,|e\rangle$ are the respective ground and excited states.
To apply the input-output formalism (\ref{io3})-(\ref{io1}), we map the Hamiltonian of the two-level atom (\ref{Hatom})
 to the Hamiltonian of a Kerr-nonlinear cavity with infinite Kerr-nonlinearity strength:
\begin{equation}\label{hsys}
H_{c}=\Omega\,a^{\dag}a+\frac{\chi}{2}a^{\dag}a^{\dag}aa\,,\,\,\,\,\,\,\,\,\,\chi\rightarrow \infty\,,
\end{equation}
which can be diagonalized as $H_c|n\rangle =E_n|n\rangle$ for $n\geq 0$ and $E_n=\Omega\,n+\frac{\chi}{2}n(n-1)$.
When $\chi\rightarrow \infty$, we have $E_0=0$, $E_1=\Omega$ and $E_n\rightarrow \infty$ for all $n\geq 2$.  As a result, the cavity cannot be excited twice and thus has exactly the same behavior as the two-level atom during the few-photon transport process. 
The total excitation number operator is $N=a^{\dag}a$ satisfying $[N, H_c]=0$. For validation, we solve the system directly in Appendix A.3 without mapping to the cavity case.

We describe the  background process as a general $2\times 2$ symmetric unitary matrix
\begin{equation}\label{cmt2}
\mathbf{C}=e^{i\phi}\left[\begin{array}{cc}t & i\,r \\i\,r & t\end{array}\right]
\end{equation} 
with $|r|^2+|t|^2=1$. Here we take $t\in [0,1]$ and $r=\sqrt{1-t^2}$. It is known that any two-port system can be written in this form. 
Moreover, with (\ref{cmt2}) and additional mirror symmetry $\mathbf{d}_1=\pm\mathbf{d}_2$, we can solve $\mathbf{d}$ from the constraint (\ref{fcc1}) as
\begin{equation}\label{cmt3}
\mathbf{d}=e^{i\frac{\phi}{2}}{\cal{M}}\frac{i(1+t)-r}{\sqrt{2(1+t)}}\left[\begin{array}{c}1 \\\pm1\end{array}\right]\,.
\end{equation}
The $\pm$ sign corresponds to the case where the atom excitation is even (odd) with respect to the mirror plane,
in which case $\mathbf{d}_1=+(-)\mathbf{d}_2$. Using the constraint (\ref{fcc2}) and assuming $\Sigma$ is real, we determine ${\cal{M}}=\sqrt{\Sigma}$. 

\begin{figure}[h]
\includegraphics[width=0.5\textwidth] {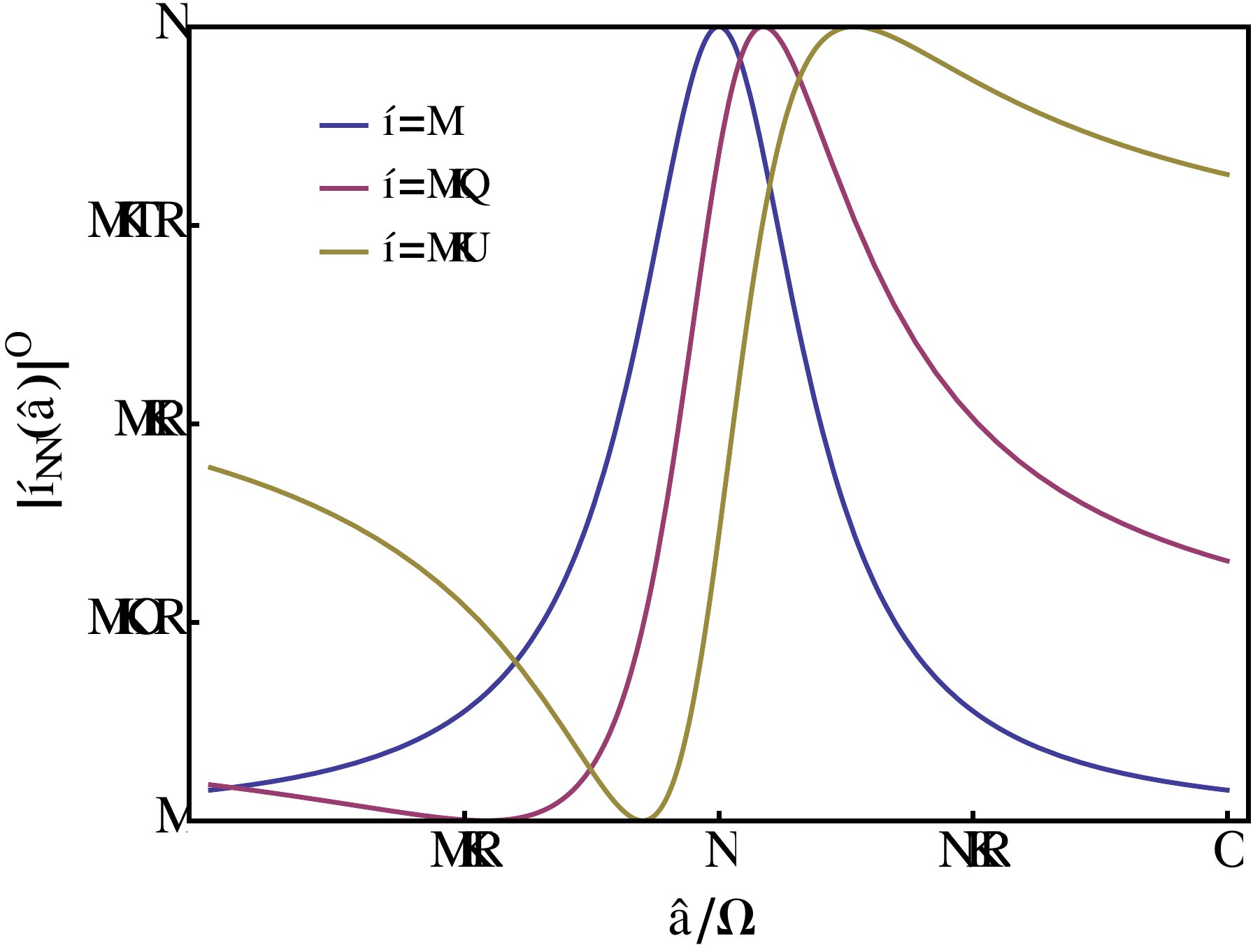} \,\,\,\,\,\,\,\,
 \caption{
The transmission coefficient as a function of photon frequency. $\Sigma=0.2\,\Omega$. Fano resonances appear when varying $t$ from $0$ to $1$. 
} \label{fignew1}
\end{figure}

Following (\ref{s1}) and (\ref{SC1}),  we compute the single photon S matrix as
\begin{equation}\label{atomS1}
S_{\mu p;\nu k}=\left[\mathbf{C}_{\mu\nu}+\mathbf{d}_{\mu}\,\mathbf{d}_{\nu}\,\frac{i}{k-\Omega+i\,\Sigma}\right]\,\delta(p-k)\equiv t_{\mu\nu}(k)\delta(p-k)\,,
\end{equation}
where $\mathbf{C}$ and $\mathbf{d}$ are shown in (\ref{cmt2}) and (\ref{cmt3}), respectively. Suppose we send in a right-moving photon with frequency $k$, according to our channel convention, $t_{11}(k)$ is the transmission amplitude and $t_{21}(k)$ is the 
reflection amplitude. In Fig.\ref{fignew1} we plot the transmission coefficient $\left|t_{11}(k)\right|^2$. 
For $t=0$ where there is no background transmission, the transmission spectrum is a Lorentizian and the maximal transmission occurs at the resonant frequency $\Omega$.
For all other cases where $t$ is between $0$ and $1$, the transmission spectrum exhibits a Fano asymmetric line shape where the transmission coefficient vary from $0$ to $1$ from
a small change in incident photon frequency.

For the two-photon transport, following (\ref{SC2}), when  $\chi\rightarrow \infty$, the connected two-photon S matrix is 
\begin{eqnarray}\label{atomSC2}
{S}^C_{\mu p_1,\nu p_2;\rho k_1,\sigma k_2}&=&\frac{i}{\pi}\mathbf{d}_{\mu}\mathbf{d}_{\nu}\mathbf{d}_{\rho}\mathbf{d}_{\sigma}
\frac{k_1+k_2-2\Omega+2i\Sigma}{\left(p_1-\Omega+i\Sigma\right)\left(p_2-\Omega+i\Sigma\right)\left(k_1-\Omega+i\Sigma\right)\left(k_2-\Omega+i\Sigma\right)}\delta(p_1+p_2-k_1-k_2)\,.
\end{eqnarray}
By (\ref{Hmi}) and (\ref{atomSC2}), if we send in two right-moving photons (in the waveguide channel $1$) with frequencies $k_1$ and $k_2$, the wavefunction of the transmitted two-photon bound state is
\begin{equation}\label{atomH}
H_{11}(x_1,x_2)=\frac{1}{\sqrt{2}\pi}\mathbf{d}_1^4\,e^{iE\frac{x_1+x_2}{2}}\,\frac{e^{-i(E/2-\Omega+i\Sigma)\left|x_1-x_2\right|}}{(k_1-\Omega+i\Sigma)(k_2-\Omega+i\Sigma)}\,,
\end{equation}
where $E\equiv k_1+k_2$ is the total frequency of the two incident photons.
Finally, the transmitted two-photon correlation function can be obtained by substituting (\ref{atomS1}) and (\ref{atomH}) into (\ref{G2y}).  As a result, we have
\begin{eqnarray}
G^{(2)}\left(0\right)=2\left|\frac{1}{\sqrt{2}\pi}t_{11}(k_1)t_{11}(k_2)+H_{11}(0,0)\right|^2\,.\label{G20}
\end{eqnarray}

\begin{figure}
\includegraphics[width=0.7\textwidth] {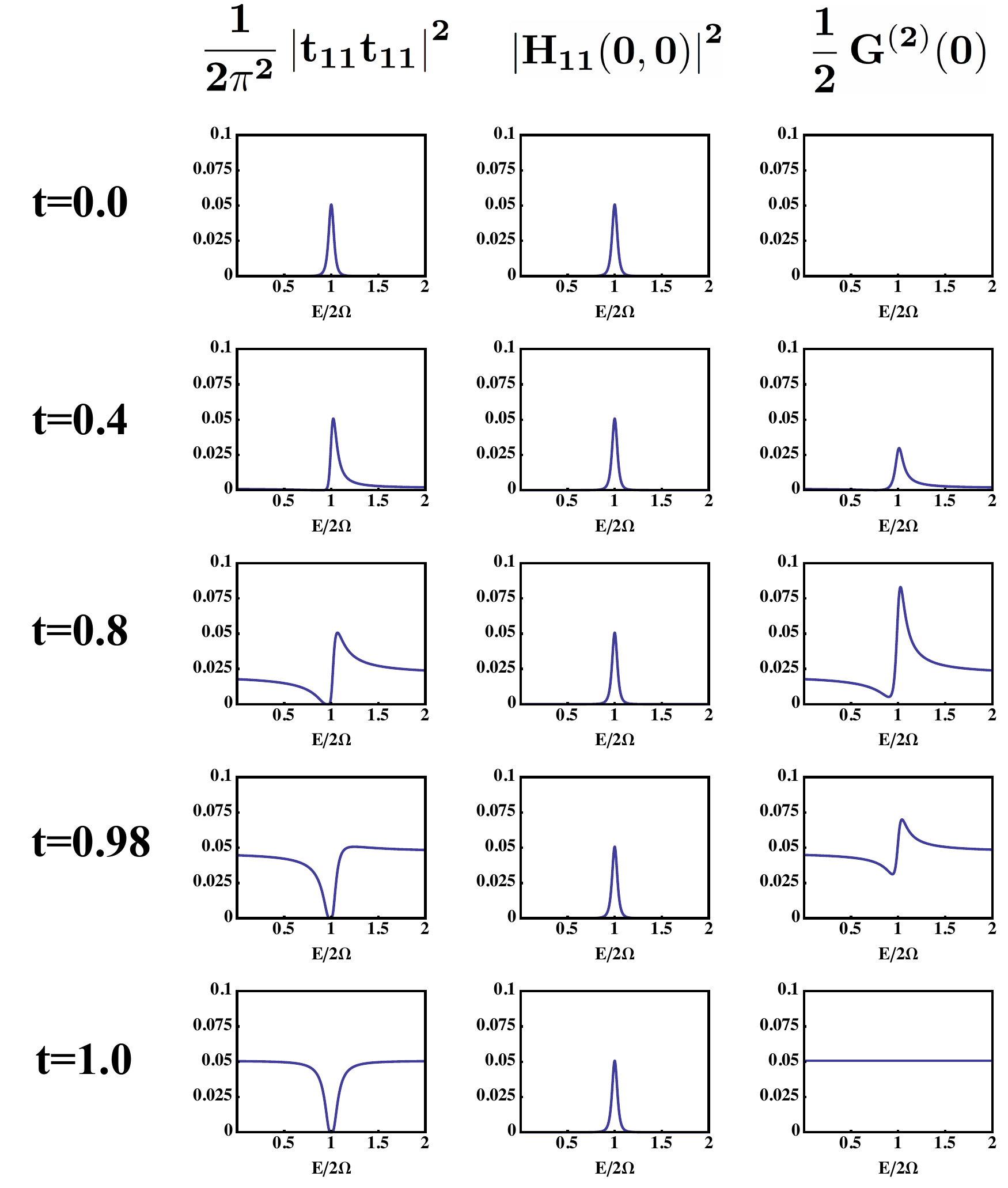} 
\caption{$\left|\frac{1}{\sqrt{2}\pi}t_{11}t_{11}\right|^2$, $\left| H_{11}(0,0)\right|^2$ and $\frac{1}{2}G^{(2)}(0)$  versus $E/2$ for different values of $t$
when  $k_1=k_2=E/2$ and $\Sigma=0.2\,\Omega$.
} \label{fig3new}
\end{figure}

From now on we focus on the case that two incident photons have the same frequencies, i.e. $k_1=k_2={E}/{2}$. In this case,
\begin{eqnarray}
\left|\frac{1}{\sqrt{2}\pi} t_{11}\,t_{11}\right|^2&=&\frac{1}{2\pi^2}\frac{\left[\left(\frac{E}{2}-\Omega\right)t+r\,\Sigma\right]^4}{\left[\left(\frac{E}{2}-\Omega\right)^2+\Sigma^2\right]^2}\label{t11t11}\\
\left|H_{11}(0,0)\right|^2&=&\frac{1}{2\pi^2}\frac{\Sigma^4}{\left[\left(\frac{E}{2}-\Omega\right)^2+\Sigma^2\right]^2}\,,\label{H11E/2}\\
\frac{1}{2}G^{(2)}\left(0\right)&=&\frac{1}{2\pi^2}\frac{\Sigma^4}{\left[\left(\frac{E}{2}-\Omega\right)^2+\Sigma^2\right]^2}\left| \left(t\,\frac{\frac{E}{2}-\Omega}{\Sigma}+r\right)^2e^{-2i\arccot t}+1\right|^2\,.\label{G11E/2}
\end{eqnarray}
In Fig.\ref{fig3new}, we plot the probability of two-photon product state 
$\frac{1}{2\pi^2}\left|t_{11}\,t_{11}\right|^2$,  the probability of the two-photon bound state $\left|H_{11}(0,0)\right|^2$ and the transmitted two-photon correlation function $\frac{1}{2}G^{(2)}(0)$ as a function of $E/2$ in three separate columns, respectively. 
In this plot, we fix $\Sigma$ and vary the scattering matrix of the direct pathway, i.e.,
the value of $t$ in (\ref{cmt2}). As $t$ increases from $0$ to $1$, $\frac{1}{2\pi^2}\left|t_{11}\,t_{11}\right|^2$ exhibit Fano asymmetric line shapes.
On the other hand, $\left|H_{11}(0,0)\right|^2$, which characterizes the contribution of the two-photon bound state and is related to the two-photon fluorescence spectrum, always
has a symmetric line shape with respect to $E$ and is independent of $t$.
Consequently, the transmitted two-photon correlation function $G^{(2)}(0)$, which has contributions from the two terms described above, also exhibit an
asymmetric line shape with respect to $E$ when $t\neq 0,1$. The physics of Fano interference therefore manifests in the two-photon correlation function.

Examining (\ref{G20}), we see that the relative phase of the two terms is also important in determining $G^{(2)}(0)$. In particular, when $t=0$, the two terms
have exactly the same magnitude but opposite phase as shown in (\ref{G11E/2}) , in which case $G^{(2)}(0)=0$ for all $E$. Thus remarkably the system exhibits complete anti-bunching for all $E$.
When $t=1$, the interference of such two terms also results in a $G^{(2)}(0)$ that is independent of $E$. In this case, $\frac{1}{2}G^{(2)}(0)=\frac{1}{2\pi^2}$ is the same as the 
$G^{(2)}(0)$ of a two-photon plane wave.

\section{VI. Summary}
To summarize, in this paper, we present a general input-output formalism for waveguide QED systems where a local cavity is coupled to multiple waveguide channels. 
We show that the parameters of this formalism are strongly constrained by general arguments such as flux conservation and time-reversal symmetry. Using this formalism, 
we study the effect of Fano interference on multi-photon transport in such systems. We show that Fano interference does manifest in the two-photon correlation function. One 
therefore can use the physics of Fano interference to tailor the quantum transport properties of multi-photon states in these systems such as tailoring the two-photon correlation 
functions by tuning the background scattering matrix.  Our work points to the importance of Fano interference in the quantum regime of light transport.

\section{ACKNOWLEDGEMENT}
This research is supported by an AFOSR-MURI program, Grant No.
FA9550-12-1-0488. 

\section{Appendix}

\subsection{A.1  A specific system that has input-output formalism (\ref{io3})-(\ref{io1})}
We provide an explicit derivation of the input-output formalism as shown in (\ref{io3})-(\ref{io1}) from the following Hamiltonian
of a specific waveguide QED system:
\begin{eqnarray}\label{H}
H=\sum_{\mu=1}^N \int dk\,k\,c_{\mu,k}^{\dag}c_{\mu,k}+
\sum_{\mu=1}^N\xi_{\mu}\int \frac{dk}{\sqrt{2\pi}}\left(c_{\mu,k}^{\dag}a+a^{\dag}c_{\mu,k}\right)+H_{\text{c}}+\sum_{i\neq j}V_{\mu\nu}\int \frac{dk}{\sqrt{2\pi}} \int \frac{dk'}{\sqrt{2\pi}}\,c_{\mu,k}^{\dag}c_{\nu,k'}\,,
\end{eqnarray}
where $c_{\mu,k} \,(c^{\dag}_{\mu,k})$ is the annihilation (creation) operator of the photon
state in the waveguide channel $\mu$ satisfying the standard commutation
relation $\left[c_{\mu,k},\, c^{\dag}_{\nu,k'}\right]=\delta(k-k')\delta_{\mu\nu}$. We consider only a narrow range
of frequencies in which the dispersion relations in all the channels can
be linearized, and all the group velocities are taken
to be 1. $H_{\text{c}}$ is the cavity Hamiltonian whose strong
nonlinearity forms the basis for
strong photon-photon interactions.
$a\, (a^{\dag})$ is the bosonic annihilation (creation) operator of the cavity that
commutes with $c_{\mu,k},\,c^{\dag}_{\nu,k}$ and satisfies the commutation relation $[a,a^{\dag}]=1$. $\xi_{\mu}$ is the coupling constant between the cavity and the corresponding waveguide channel $\mu$. The last term in (\ref{H}) is the direct coupling among waveguide channels with 
$V_{\mu\nu}$ be the coupling constant between channels $\mu$ and $\nu$. $V_{\mu\nu}$ satisfies the constraint  $V_{\mu\nu}=V_{\nu\mu}$ as required by the Hermitian of the Hamiltonian. 

For the Hamiltonian (\ref{H}), the Heisenberg equations are
\begin{eqnarray}
\frac{d}{dt}c_{\mu,k}&=&-i\,k\,c_{\mu,k}-i\,\frac{\xi_{\mu}}{\sqrt{2\pi}}\,a-i\sum_{\nu\neq \mu}\frac{V_{\mu\nu}}{\sqrt{2\pi}}\int \frac{dk'}{\sqrt{2\pi}}\,c_{\nu,k'}\,,\label{Heisen1}\\
\frac{d}{dt}a &=&-i\,\left[a,\,H_{\text{c}}\right] -i\,\sum_{\mu=1}^N\xi_{\mu}\int \frac{dk}{\sqrt{2\pi}} c_{\mu,k}\,\label{Heisen2}\,. 
\end{eqnarray}

We define
\begin{equation}
\Phi_{\mu}(t)\equiv\int\,\frac{dk}{\sqrt{2\pi}}\, c_{\mu,k}(t)\,,
\end{equation}
and define the input and output operators 
\begin{eqnarray}\label{cinout}
c_{\mu,\text{in}}(t)&=&\int\,\frac{dk}{\sqrt{2\pi}}\, c_{\mu,k}(t_0)\,e^{-ik(t-t_0)}\,,\nonumber\\
c_{\mu,\text{out}}(t)&=&\int\,\frac{dk}{\sqrt{2\pi}}\, c_{\mu,k}(t_1)\,e^{-ik(t-t_1)}\,,
\end{eqnarray}
with $t_0\rightarrow -\infty\,,t_1\rightarrow +\infty$.

After multiplying (\ref{Heisen1}) by  the factor $\exp(ikt)$, we integrate it from an initial time $t_0<t$ to get
\begin{equation}\label{in}
c_{\mu,k}(t)=c_{\mu,k}(t_0)e^{-ik(t-t_0)}-i\frac{\xi_{\mu}}{\sqrt{2\pi}}\int_{t_0}^td\tau\, a(\tau)e^{-ik(t-\tau)}-i\sum_{\nu\neq \mu}\frac{V_{\mu\nu}}{\sqrt{2\pi}}\int_{t_0}^td\tau\, \Phi_{\nu}(\tau)e^{-ik(t-\tau)}\,.
\end{equation}
Integrating (\ref{in}) with respect to $k$, we get
\begin{equation}\label{in1}
\Phi_{\mu}(t)=c_{\mu,\text{in}}(t)-i\frac{\xi_{\mu}}{2}a(t)-i\sum_{\nu\neq \mu}\frac{V_{\mu\nu}}{2}\Phi_{\nu}(t)\,.
\end{equation}

Similarly, we integrate (\ref{Heisen1}) up to a final time $t_1>t$ and obtain 
\begin{equation}\label{out}
\Phi_{\mu}(t)=c_{\mu,\text{out}}(t)+i\frac{\xi_{\mu}}{2}a(t)+i\sum_{\nu\neq \mu}\frac{V_{\mu\nu}}{2}\Phi_{\nu}(t)\,.
\end{equation}

We introduce matrices
\begin{eqnarray}\label{cmat}
\mathbf{ c}_{\text{in}}(t)=\left[\begin{array}{c} c_{1,\text{in}}(t)\\c_{2,\text{in}}(t)\\ \vdots \\c_{N,\text{in}}(t)\end{array}\right]\,,\,\,\,\,
\mathbf{ c}_{\text{out}}(t)=\left[\begin{array}{c} c_{1,\text{out}}(t)\\c_{2,\text{out}}(t)\\ \vdots \\c_{N,\text{out}}(t)\end{array}\right]\,,\,\,\,\,
\boldsymbol{\xi}=\left[\begin{array}{c} \xi_1\\\xi_2\\ \vdots \\\xi_N\end{array}\right]\,,\,\,\,\,
\mathbf{V}=\left[\begin{array}{cccc}0 & V_{12} & \cdots & V_{1N} \\V_{21} & 0 & \cdots & V_{2N} \\\vdots & \vdots & \ddots & \vdots \\V_{N1} & V_{N2} & \cdots & 0\end{array}\right]\,.
\end{eqnarray}
Then (\ref{in1}), (\ref{out}) and (\ref{Heisen2}) can be written as:
\begin{eqnarray}
\left(\mathbf{ I}+\frac{i}{2}\mathbf{ V}\right)\,\mathbf{ \Phi}(t)&=&\mathbf{ c}_{\text{in}}(t)-\frac{i}{2}\,a(t)\,\boldsymbol{\xi}\,,\\
\left(\mathbf{ I}-\frac{i}{2}\mathbf{ V}\right)\,\mathbf{ \Phi}(t)&=&\mathbf{ c}_{\text{out}}(t)+\frac{i}{2}\,a(t)\,\boldsymbol{\xi}\,,\\
\frac{d}{dt}a &=&-i\left[a, H_{\text{c}}\right]-i\,\boldsymbol{\xi}^T\,\mathbf{\Phi}(t)\,.
\end{eqnarray}

Eliminating the variable $\mathbf{\Phi}$ leads to the input-output formalism:
\begin{eqnarray}\label{spio}
\mathbf{ c}_{\text{out}}(t)&=&\left(\mathbf{ I}-\frac{i}{2}\mathbf{ V}\right)\left(\mathbf{ I}+\frac{i}{2}\mathbf{ V}\right)^{-1}\mathbf{ c}_{\text{in}}(t)
-{i}\,a(t)\left(\mathbf{ I}+\frac{i}{2}\mathbf{ V}\right)^{-1}\boldsymbol{\xi}\nonumber\\
\mathbf{ c}_{\text{in}}(t)&=&\left(\mathbf{ I}+\frac{i}{2}\mathbf{ V}\right)\left(\mathbf{ I}-\frac{i}{2}\mathbf{ V}\right)^{-1}\mathbf{ c}_{\text{out}}(t)
+{i}\,a(t)\left(\mathbf{ I}-\frac{i}{2}\mathbf{ V}\right)^{-1}\boldsymbol{\xi}\nonumber\\
\frac{d}{dt}a(t)&=&-i\,\left[a,\,H_{\text{c}}\right](t)-\frac{1}{2}\boldsymbol{\xi}^T\left(\mathbf{ I}+\frac{i}{2}\mathbf{ V}\right)^{-1}\boldsymbol{\xi}\,a(t)-i\,\boldsymbol{\xi}^T\left(\mathbf{ I}+\frac{i}{2}\mathbf{ V}\right)^{-1}\mathbf{ c}_{\text{in}}(t)
\end{eqnarray} 

Finally, by identifying 
\begin{eqnarray}\label{V}
&&\mathbf{C}\equiv\left(\mathbf{ I}-\frac{i}{2}\mathbf{ V}\right)\left(\mathbf{ I}+\frac{i}{2}\mathbf{ V}\right)^{-1}\,,\,\,\,\,\,
\mathbf{d}\equiv-i\,\left(\mathbf{ I}+\frac{i}{2}\mathbf{ V}\right)^{-1}\boldsymbol{\xi}\,,\nonumber\\
&&\Sigma\equiv\frac{1}{2}\,\boldsymbol{\xi}^T\left(\mathbf{ I}+\frac{i}{2}\mathbf{ V}\right)^{-1}\boldsymbol{\xi}\,,\,\,\,\,\,\,\,\,\,\,\,\,\,\,\,\,\,\,\boldsymbol{\kappa}\equiv-i\,\left(\mathbf{ I}+\frac{i}{2}\mathbf{ V}^T\right)^{-1}\boldsymbol{\xi}\,,
\end{eqnarray}
we reduce (\ref{spio}) to the general form (\ref{io3})-(\ref{io1}). One can check explicitly that $\mathbf{C}$, $\mathbf{d}$, $\boldsymbol{\kappa}$, and $\Sigma$ defined in (\ref{V}) satisfy the general constraints (\ref{fluxc}) and (\ref{trs}). 

The Hamiltonian (\ref{H}) has the parity-time symmetry.
Let $P$ and $T$ be the respective parity and (antiunitary) time-reversal operator such that 
\begin{equation}
P\, c_{\mu}(x)\, P^{-1} =c_{\mu}(-x)\,,\,\,\,\,\,\,T\, c_{\mu}(x)\, T^{-1}=c_{\mu}(x)\,,\,\,\,\,\,\,P\, a\, P^{-1} =a\,,\,\,\,\,\,\,T\, a\, T^{-1} =a\,,
\end{equation}
where $c_{\mu}(x)$ is the annihilation operator in the real operator. Since $c_{\mu, k} = \int \frac{dx}{\sqrt{2\pi}} c_{\mu}(x) e^{-ikx}$, we have
\begin{equation}
P\, c_{\mu,k}\, P^{-1} =c_{\mu,-k}\,,\,\,\,\,\,\,T\, c_{\mu,k}\, T^{-1}=c_{\mu,-k}\,,\,\,\,\,\,\,P\, a\, P^{-1} =a\,,\,\,\,\,\,\,T\, a\, T^{-1} =a\,.
\end{equation}
Let $\Theta \equiv PT$, for the Hamiltonian (\ref{H}), one can check that $\left[\Theta, H\right]=0$. Note that $\Theta$ is antiunitary, 
applying  $\Theta$ to the Heisenberg operator $a(t)$ and $c_{\mu,\text{in}}(t_0)$ gives
\begin{eqnarray}
\Theta \,a(t)\, \Theta^{-1} &=& \Theta \left(e^{iHt} a e^{-iHt}\right) \Theta^{-1} = e^{-iHt} \left(\Theta\, a\, \Theta^{-1}\right) e^{iHt}=e^{-iHt} a e^{iHt} = a(-t)\,,\nonumber\\
\Theta\,{c}_{\mu,k}(t_0)\, \Theta^{-1}&=& \Theta \left(e^{iHt_0} {c}_{\mu,k} e^{-iHt_0}\right) \Theta^{-1} = e^{-iHt_0} \left(\Theta\, {c}_{\mu,k}\, \Theta^{-1}\right) e^{iHt_0}
=e^{-iHt_0} {c}_{\mu,k} e^{iHt_0} ={c}_{\mu,k}(-t_0) = {c}_{\mu,k}(t_1)\,.\nonumber
\end{eqnarray}
As a result, for the  input and output operators (\ref{cinout}), we have
\begin{eqnarray}
\Theta\,c_{\mu,\text{in}}(t)\,\Theta^{-1}=
\int\,\frac{dk}{\sqrt{2\pi}}\, \Theta \,c_{\mu,k}(t_0)\Theta^{-1}\,e^{ik(t-t_0)}= \int\,\frac{dk}{\sqrt{2\pi}}\,c_{\mu,k}(t_1)\,e^{-ik(-t-t_1)} = c_{\mu,\text{out}}(-t)\,,
\end{eqnarray}
which agrees with (\ref{pts}).

\subsection{A.2. Effective Hamiltonian}
In this Appendix, we prove that the Green functions in (\ref{SC1}) and (\ref{SC2})  can be computed using the effective Hamiltonian of the cavity
\begin{equation}\label{effH1}
H_{\text{eff}}\equiv H_{\text{c}}-i\,\Sigma\, a^{\dag}a
\end{equation}
without involving any waveguide degrees of freedom. 

We first prove that the propagator of the cavity can be computed using the effective Hamiltonian. That is, when $H_{\text{c}}=H_{\text{c}}^{(0)}\equiv\omega\, a^{\dag}a$,  
\begin{equation} \label{id}
G^{(0)}({t', t})=\widetilde{G}^{(0)}({t',  t})\,,
\end{equation}
where
\begin{eqnarray}
G^{(0)}({t', t})&\equiv&\langle 0| {\cal{T}}\,{a}(t') {a^{\dag}}(t) |0\rangle\,,\\
\widetilde{G}^{(0)}({t',  t})&\equiv&\langle 0| {\cal{T}}\,\widetilde{a}(t') \widetilde{a^{\dag}}(t) |0\rangle\,. \label{G0t}
\end{eqnarray}
$a(t)$ and $a^{\dag}(t)$ are Heisenberg operators in the input-output formalism (\ref{io3})-(\ref{io1}).  $\widetilde{a}(t)$ and $\widetilde{a^{\dag}}(t)$ are evolved by the effective 
Hamiltonian (\ref{effH1}) as
\begin{equation}\label{effa}  
\widetilde{a}(t)\equiv e^{i H_{\text{eff}} t}\, a\, e^{-i H_{\text{eff}} t}\,,\,\,\,\,\,\,\,\,\widetilde{a^{\dag}}(t)\equiv e^{i H_{\text{eff}} t} \,a^{\dag}\, e^{-i H_{\text{eff}} t}\,.
\end{equation}
With the identity (\ref{id}), the computation of the propagator is simplified since no operators of waveguide photons are involved in (\ref{G0t}).  
We only need to solve a system which has a finite, and typically small, number of degrees of freedom. 

The proof is as follows. When $t'>t$,
\begin{eqnarray}
\frac{\partial}{\partial t'} G^{(0)}({t', t})&=& \langle 0|\,\frac{d{a}(t')}{dt'}\, {a^{\dag}}(t)\, |0\rangle\nonumber\\
&=&-i\,(\omega-i\,\Sigma)\, \langle 0|{a}(t')\, {a^{\dag}}(t)|0\rangle+\langle 0|\boldsymbol{\kappa}^T\,\mathbf{ c}_{\text{in}}(t')a^{\dag}(t)|0\rangle\label{eff11}\\
&=&-i\,(\omega-i\,\Sigma)\, G^{(0)}({t', t})\,,\label{eff12}\\
\frac{\partial}{\partial t} G^{(0)}({t', t})&=& \langle 0|\,{a}(t')\frac{d{a}^{\dag}(t)}{dt} \,|0\rangle\nonumber\\
&=&i\,(\omega+i\,\Sigma^*+i\,\mathbf{d}^{\dag}\,\mathbf{C}\,\boldsymbol{\kappa}^*)\, \langle 0|{a}(t')\, {a^{\dag}}(t)|0\rangle+\langle 0|a(t')\,\mathbf{c}^{\dag}_{\text{out}}(t)\,\mathbf{C}\,\boldsymbol{\kappa}^*|0\rangle\label{eff21}\\
&=&i\,(\omega-i\,\Sigma)\, G^{(0)}({t', t})\,.\label{eff22}
\end{eqnarray}
In (\ref{eff11}) and (\ref{eff21}), we use the input-output formalism (\ref{io3})-(\ref{io1}). To obtain (\ref{eff12}) and (\ref{eff22}), we use the respective quantum casualties (\ref{qc1}) and (\ref{qc2}) so that 
$\langle 0|\mathbf{ c}_{\text{in}}(t')a^{\dag}(t)|0\rangle=\langle 0|a^{\dag}(t)\,\mathbf{ c}_{\text{in}}(t')|0\rangle=0$ and 
$\langle 0|a(t')\,\mathbf{c}^{\dag}_{\text{out}}(t)|0\rangle=\langle 0|\mathbf{c}^{\dag}_{\text{out}}(t)\,a(t')|0\rangle=0$. In (\ref{eff22}), we also use the constraint of flux conservation (\ref{fluxc}) to transform $\Sigma^*$ to $\Sigma$.
On the other hand, by (\ref{effa}), one can compute 
\begin{eqnarray}
\frac{\partial}{\partial t'} \widetilde{G}^{(0)}({t', t})&=& \langle 0|\,\frac{d\widetilde{a}(t')}{dt'}\, \widetilde{a^{\dag}}(t)\, |0\rangle=-i\,\langle 0|\,[\widetilde{a}, H_{\text{eff}}](t')\, \widetilde{a^{\dag}}(t)\, |0\rangle \nonumber\\
&=&-i\,(\omega-i\,\Sigma)\, \langle 0|\widetilde{a}(t')\, \widetilde{a^{\dag}}(t)|0\rangle=-i\,(\omega-i\,\Sigma)\, \widetilde{G}^{(0)}({t', t})\,,\label{eff31}\\
\frac{\partial}{\partial t} \widetilde{G}^{(0)}({t', t})&=& \langle 0|\widetilde{a}(t')\frac{d\widetilde{a}^{\dag}(t)}{dt} \,|0\rangle=-i\,\langle 0|\,\widetilde{a}(t')\,[\widetilde{a^{\dag}}, H_{\text{eff}}](t)\,  |0\rangle\nonumber\\
&=&i\,(\omega-i\,\Sigma)\, \langle 0|\widetilde{a}(t')\, \widetilde{a^{\dag}}(t)|0\rangle=i\,(\omega-i\,\Sigma)\, \widetilde{G}^{(0)}({t', t})\,.\label{eff32}
\end{eqnarray}
So $G^{(0)}({t', t})$ and $\widetilde{G}^{(0)}({t', t})$ satisfy exactly the same differential equations when $t'>t$. They also have the same initial values at $t'= t$. Therefore, by the uniqueness theorem for differential equations, we complete the proof of  (\ref{id}). 

Now for a general Hamiltonian $H_{\text{c}}=H_{\text{c}}^{(0)}+V$, according to the perturbation theory in quantum field theory, all Green functions in principle are completely determined by the propagator and the interaction vertices. The vertices only reply 
on the form of the interaction term $V$ and is independent of the waveguide photons.   As a result, all Green functions, including higher-order ones, can be computed by the effective Hamiltonian (\ref{effa}). 

In Ref.\cite{sfs,xf}, the effective Hamiltonian is obtained in the path integral formalism by integrating out the waveguide degrees of freedom in the full Hamiltonian.
The derivation here only relies on the input-output formalism (\ref{io3})-(\ref{io1}) and the resulting quantum causalities (\ref{qc1})-(\ref{qc2}) .

\subsection{A.3.  A two-mode waveguide coupled to a two-level atom}
Following Ref.\cite{fks}, here we present the input-output formalism for the
 waveguide QED systems that consist of a two-level atom coupled to $N$ waveguide channels:
\begin{eqnarray}
\frac{d}{dt}\,\sigma_-&=&-i\,\Omega\,\sigma_--\,\Sigma\,\sigma_--\sigma_z\,\boldsymbol{\kappa}^T\,\mathbf{ c}_{\text{in}}\,\label{2io3}\\
\mathbf{ c}_{\text{out}}(t)&=&\mathbf{C}\,\mathbf{ c}_{\text{in}}(t)+\,\sigma_-(t)\,\mathbf{d}\,.\label{2io1}
\end{eqnarray}
In the equations above, $\Omega$ is the atomic transition frequency.
$\sigma_{+}\,(\sigma_-)$ is the atomic raising (lowering) operator. These operators satisfy the commutation relations $\left[\sigma_z,\,\sigma_{\pm}\right]=\pm 2\,\sigma_{\pm}$ and 
$\left[\sigma_+,\,\sigma_-\right]=\sigma_z$. The total excitation number here is 
$N=\frac{1}{2}\left(\sigma_z+1\right)=\sigma_+\sigma_-$. With (\ref{2io3}), one can show that
\begin{eqnarray}
\frac{d}{dt}N=-\left(\Sigma +\Sigma^*\right)\,N + \sigma_+\,\boldsymbol{\kappa}^T\mathbf{ c}_{\text{in}}+\mathbf{ c}^{\dag}_{\text{in}}\boldsymbol{\kappa}^*\sigma_-\,.\label{2io2}
\end{eqnarray}

We first sketch the proof that the constraints imposed by flux conservation, quantum causality and time-reversal symmetry
are the same as that in the cavity case. Applying the input-output formalism (\ref{2io1})-(\ref{2io2}) to the flux conservation condition
(\ref{fluxio}) gives the same constraint as that in (\ref{fluxc}). Similarly, we define the (antiunitary) time-reversal operator $\Theta$ as
\begin{equation}\label{pts2atom}
\Theta\, \mathbf{c}_{\text{in}}(t)\, \Theta^{-1}=\mathbf{c}_{\text{out}}(-t)\,,
\,\,\,\,\,\,\,\Theta \,\sigma_{\pm}(t) \,\Theta^{-1}=\sigma_{\pm}(-t)\,.
\end{equation} 
Then the invariance of the input-output formalism (\ref{2io3})-(\ref{2io1}) under the time-reversal operation (\ref{pts2atom}) gives the same constraint as that in (\ref{trs}).
Only the quantum causality condition requires a slightly different treatment. From the input-output formalism (\ref{2io3})-(\ref{2io1}), we can prove the following quantum causality condition
\begin{eqnarray}
\left[\sigma_-(t)\,,\mathbf{c}^{\dag}_{\text{in}}(t')\right]&=&-\mathbf{d}^{\dag}\mathbf{C}\,\left[\sigma_-(t)\,,\sigma_+(t')\right]\,\theta(t-t')\,,\label{qc1atom}\\
\left[\sigma_-(t)\,,\mathbf{c}^{\dag}_{\text{out}}(t')\right]&=&\mathbf{d}^{\dag}\,\left[\sigma_-(t)\,,\sigma_+(t')\right]\,\theta(t'-t)\,.\label{qc2}\label{qc2atom}
\end{eqnarray}
With (\ref{qc1atom}) and the input-output formalism (\ref{2io3}), we can compute
\begin{eqnarray}
\frac{d}{dt}N&=&\frac{1}{2}\frac{d}{dt}\left[\sigma_+(t), \sigma_-(t)\right]= \frac{1}{2}\left[\frac{d}{dt}\sigma_+,\,\sigma_-\right]+\frac{1}{2}\left[\sigma_+,\,\frac{d}{dt}\sigma_-\right]\nonumber\\
&=&-\left(\Sigma+\Sigma^*\right) \left(N-\frac{1}{2}\right)+\frac{1}{4}\left(\boldsymbol{\kappa}^T\,\mathbf{C}^{\dag}\,\mathbf{d}\,+\,\mathbf{d}^{\dag}\,\mathbf{C}\,\boldsymbol{\kappa}^*\right)
+ \sigma_+\,\boldsymbol{\kappa}^T\mathbf{ c}_{\text{in}}+\mathbf{ c}^{\dag}_{\text{in}}\boldsymbol{\kappa}^*\sigma_-\,.\label{anotherN}
\end{eqnarray}
By comparing (\ref{2io2}) with (\ref{anotherN}), we obtain the same constraint as that in (\ref{constr1}). Therefore, all the constraints for the two-level atom are exactly the same as that in the cavity case.
As a result, for the system consisting of a two-mode waveguide coupled to a two-level atom, the parameters in the input-output formalism (\ref{2io3})-(\ref{2io2}) 
have the same form as described in (\ref{cmt2}) and (\ref{cmt3}).

Now we can compute the single photon S matrix $S_{\mu p;\nu k}$ and two-photon S matrix $S_{\mu p_1,\nu p_2;\rho k_1, \sigma k_2}$ using the method presented in \cite{fks}. 
The computations in \cite{fks} only relies on the input-output formalism and can be straightforwardly generalized to the multi-channel case here. Following \cite{fks}, one can check that the final results of the S matrices are exactly the same as that in 
(\ref{atomS1}) and (\ref{atomSC2}).

Finally, we point out that the input-output formalism for a two-level atom (\ref{2io3})-(\ref{2io1}) can also be adopted to develop
 an input-output formalism for a free fermion coupled to waveguide channels. By applying the single-site Jordan-Wigner transformation
\begin{eqnarray}
\sigma_+ = f^{\dag}\,,\,\,\,\,\,\,\sigma_-=f\,,\,\,\,\,\,\,\sigma_z=2f^{\dag}f-1
\end{eqnarray}
to the input-output formalism (\ref{2io3})-(\ref{2io1}), we have 
\begin{eqnarray}
\frac{d}{dt}\,f&=&-i\,\Omega\,f-\,\Sigma\,f-(2f^{\dag}f-1)\,\boldsymbol{\kappa}^T\,\mathbf{ c}_{\text{in}}\,\label{3io3}\\
\mathbf{ c}_{\text{out}}(t)&=&\mathbf{C}\,\mathbf{ c}_{\text{in}}(t)+\,f(t)\,\mathbf{d}\,,\label{3io1}
\end{eqnarray}
where $f,f^{\dag}$ are the fermionic annihilation and creation operators satisfying $\left\{f,f^{\dag}\right\}=1$. For the single-channel case, we can check that (\ref{3io3})-(\ref{3io1}) agrees with 
the input-output formalism derived from the Hamiltonian 
\begin{eqnarray}
H_f = \int dk\,k\,c_k^{\dag}c_k +\Omega\, f^{\dag}f+ \sqrt{\frac{\gamma}{2\pi}}\int dk\left(f^{\dag}c_k+c_k^{\dag}f\right)\,.\nonumber
\end{eqnarray}

In sum, we show that the Kerr-nonlinear cavity with infinite Kerr-nonlinearity strength, the two-level atom and the free fermion have the same S matrix in the few-photon scattering process since they
have the same internal energy levels.

\end{document}